\documentclass[aps,prd,twocolumn,superscriptaddress,showpacs,nofootinbib]{revtex4}

\usepackage{graphicx} \usepackage{color} \usepackage{natbib}

\usepackage{amsmath,amssymb,amsfonts}
\usepackage{epsfig}
\usepackage{longtable}
\usepackage{color}

\begin{document}

%\preprint{}

\title{Revealing the high-density equation of state through binary neutron star mergers}


%\author{}
\author{A.~Bauswein}
%\thanks{}
\affiliation{Department of Physics, Aristotle University of
  Thessaloniki, GR-54124 Thessaloniki, Greece}
\author{N.~Stergioulas}
\affiliation{Department of Physics, Aristotle University of
  Thessaloniki, GR-54124 Thessaloniki, Greece}
\author{H.-T.~Janka}
\affiliation{Max-Planck-Institut f\"ur
  Astrophysik, Karl-Schwarzschild-Str.~1, 85748 Garching, Germany}


\date{\today}

\begin{abstract}
We present a novel method for revealing the equation of state of high-density neutron star matter through gravitational waves emitted during the postmerger phase of a binary neutron star system. The method relies on a small number of detections of the peak frequency in the postmerger phase for binaries of different (relatively low) masses, in the most likely range of expected detections. From such observations, one can construct the derivative of the peak frequency versus the binary mass, in this mass range. Through a detailed study of binary neutron star mergers for a large sample of equations of state, we show that one can extrapolate the above information to the highest possible mass (the threshold mass for black hole formation in a binary neutron star merger). In turn, this allows for an empirical determination of the maximum mass of cold, nonrotating neutron stars to within~$0.1 M_{\odot  }$, while the corresponding radius is determined to within a few percent. Combining this with the determination of the radius of cold, nonrotating neutron stars of $1.6~M_{\odot}$ (to within a few percent, as was demonstrated in Bauswein et al., PRD, 86, 063001, 2012), allows for a clear distinction of a particular candidate equation of state among a large set of other candidates. Our method is particularly appealing because it reveals simultaneously the moderate and very high-density parts of the equation of state, enabling the distinction of mass-radius relations even if they are similar at typical neutron star masses. Furthermore, our method also allows to deduce the maximum central energy density and maximum central rest-mass density of cold, nonrotating neutron stars with an accuracy of a few per cent.
\end{abstract}

%\keywords{}

\maketitle

%for the analysis
% important matlab scripts: mostly using bigtable.dat
% 
% emaxfstab.m
% fstabMstab.m
% mthresemax.m
% mthresrhomax.m
% rhomaxfstab.m
% rmaxfstab.m
% slopefpeak.m
% 
% plotfpeakmtot.m

%EoS equation of state
%GW gravitational wave
%NS neutron star
%BH black hole
%DRO differentially rotating object

\section{Introduction}
The Advanced LIGO~\cite{2010CQGra..27h4006H} and Advanced Virgo~\cite{2006CQGra..23S.635A} gravitational-wave detectors are expected to observe between 0.4 and 400 mergers of binary neutron stars (NSs) per year, when they start operating at their design sensitivity \cite{2010CQGra..27q3001A}.\footnote{Similar rates are estimated for the upcoming KAGRA instrument~\cite{2010CQGra..27h4004K}.} The Einstein Telescope~design \cite{2010CQGra..27a5003H} promises roughly \(10^{3 }\) times higher detection rates. The merger of NSs is a consequence of gravitational wave (GW) emission, which extracts energy and angular momentum from the binary and thus forces the binary components on inspiraling trajectories. Events within a few tens of Mpc are particulary interesting, because they bear the potential to constrain the (still largely unknown)  equation of state (EoS) of neutron-star matter (see~\cite{2010CQGra..27k4002D,2011GReGr..43..409A,BaumgarteShapiro,2012LRR....15....8F,Rezzolla} for reviews and e.g.~\cite{2007PhR...442..109L,2012ARNPS..62..485L} for a discussion of the current EoS and NS constraints). The properties of cold, high-density matter are encoded in the stellar properties of nonrotating NSs, since the EoS uniquely defines the stellar structure via the Tolman-Oppenheimer-Volkoff equations~\cite{1939PhRv...55..364T,1939PhRv...55..374O}. Since the dynamics of a merger is crucially affected by the properties of NSs, the GW signal carries information on the binary parameters and the EoS (e.g.~\cite{1994PhRvD..50.6247Z,1996A&A...311..532R,2005PhRvL..94t1101S,2005PhRvD..71h4021S,2007A&A...467..395O,2007PhRvL..99l1102O,2008PhRvD..77b4006A,2008PhRvD..78b4012L,2008PhRvD..78h4033B,2009PhRvD..80f4037K,2011MNRAS.418..427S,2011PhRvD..83d4014G,2011PhRvD..83l4008H,2011PhRvL.107e1102S,2012PhRvL.108a1101B,2012PhRvD..86f3001B,2013MNRAS.430.2585R,2013PhRvL.111m1101B,2013PhRvD..88d4026H,2013arXiv1311.4443B,2014MNRAS.437L..46R,2014arXiv1402.6244B,2014arXiv1403.5672T}).

For sufficiently nearby events, the chirp-like inspiral GW signal reveals the total binary mass and the mass ratio of the merging NSs (e.g.~\cite{1993PhRvD..47.2198F,1994PhRvD..49.2658C,2005PhRvD..71h4008A,2008CQGra..25r4011V,2013ApJ...766L..14H,2013arXiv1304.1775T}). During the late inspiral phase, deviations from the point-particle behavior may be used to determine stellar properties of the inspiraling NSs (NS radii or the NS moment of inertia) with some accuracy (e.g.~\cite{2002PhRvL..89w1102F,2008PhRvD..77b1502F,2009PhRvD..79l4032R,2010PhRvD..81l3016H,2012PhRvD..85l3007D,2013PhRvL.111g1101D,2013PhRvD..88b3009Y,2013PhRvD..88j4040M,2013PhRvD..88d4042R,2013arXiv1310.8288F,2013arXiv1310.8358Y,2014arXiv1402.5156W}). As an additional method, one may detect the dominant oscillations of the postmerger remnant, which (unless there is prompt collapse to a black hole (BH)) is a hot, massive, differentially rotating NS (which is observationally the most likely case)~\cite{1994PhRvD..50.6247Z,1996A&A...311..532R,2000ApJ...528L..29B,2003ApJ...583..410L,2005PhRvL..94t1101S,2005PhRvD..71h4021S,2007A&A...467..395O,2007PhRvL..99l1102O,2008PhRvD..77b4006A,2008PhRvD..78b4012L,2008PhRvD..78h4033B,2009PhRvD..80f4037K,2011MNRAS.418..427S,2011PhRvD..83d4014G,2011PhRvD..83l4008H,2011PhRvL.107e1102S,2012PhRvL.108a1101B,2012PhRvD..86f3001B,2012PhRvD..86f4032P,2013MNRAS.430.2585R,2013PhRvL.111m1101B,2013PhRvD..88f4009G,2013PhRvD..88d4026H,2013arXiv1306.4034K,2013arXiv1311.4443B,2014MNRAS.437L..46R,2014arXiv1403.3680N,2014arXiv1403.5672T}. The dominant peak in the gravitational wave spectrum of the postmerger phase originates from a   fundamental quadrupolar (\(m=2) \) fluid oscillation mode (see \cite{2011MNRAS.418..427S} for an extraction of the mode pattern, which confirms this description), 
which appears as a pronounced peak in the GW 
spectrum, in the range between \(2-3.5 \)~kHz. Recently, it was found that for binaries with a total mass of 
about $2.7~M_{\odot}$ the frequency of this peak determines the radius of a cold, nonrotating NS with a mass of $1.6~M_{\odot}$ to within a few percent \cite{2012PhRvL.108a1101B,2012PhRvD..86f3001B}~\footnote{Note that the radii of NSs with masses somewhat different than 1.6~$M_{\odot}$ are also obtained with good accuracy.}, which was confirmed in~\cite{2013PhRvD..88d4026H}. Even a single such detection would thus tightly constrain the EoS in the density range of  $1.6~M_{\odot}$. Observations of more massive binaries would provide estimates for the radii of more massive nonrotating NSs, since they probe a higher density regime~\cite{2012PhRvD..86f3001B}.

The detection of binary NS mergers with masses larger than  $2.7~M_{\odot}$ is particularly interesting, because the determination of the threshold binary mass to BH collapse sets a tight constraint on the maximum mass of cold, nonrotating NSs, as was shown recently in~\cite{2013PhRvL.111m1101B} (notice that current pulsar observations provide a lower limit to the maximum mass of about 2~$M_{\odot}$~\cite{2010Natur.467.1081D,Antoniadis26042013}). For a given EoS the threshold binary mass to BH collapse depends in a clear way on the maximum mass of cold, nonrotating NSs and on the radius of a star 
with 1.6~$M_{\odot}$~\cite{2013PhRvL.111m1101B}. Thus, given an estimate for $R_{1.6}$ (e.g. from the inspiral GW signal or from the postmerger GW peak frequency) the determination of the threshold mass to BH collapse yields a constraint on the maximum mass of cold, nonrotating NSs.  

For most EoSs the threshold mass to BH collapse  is in the range of $3-4~M_{\odot}$~\cite{2013PhRvL.111m1101B}. This implies a serious obstacle for \textit{directly} determining the threshold mass, if NS mergers are taking place more frequently with a lower total binary mass of about $2.7 M_{\odot}$ (as is suggested by the mass distribution of observed NS binaries, see \cite{2012ARNPS..62..485L} for a compilation, and by theoretical population synthesis studies, see e.g.~\cite{2012ApJ...759...52D}). Moreover, for binary masses very near to the threshold mass for prompt collapse to a BH, the duration of the postmerger signal becomes shorter, decreasing further the expected detection rate.

In this work we show that the detection of the postmerger GW emission of two low-mass NS binary mergers with slightly different masses can be employed to estimate the threshold mass. Thus, the binary systems that are most likely to be detected, may reveal the threshold mass to BH collapse and, in turn, the maximum mass of cold, nonrotating NSs (to within \(0.1~M_{\odot}\)).  The corresponding radius is determined to within a few percent. Combining this with the determination of the radius of cold, nonrotating neutron stars of~$1.6~M_{\odot}$~\cite{2012PhRvD..86f3001B}, allows for a clear distinction of a particular candidate equation of state among a large set of other candidates.

In this paper NS masses refer to the \textit{gravitational mass in isolation}, and \textit{binary masses are reported as the sum of the gravitational masses in isolation} of the individual binary components. We use the term ``low-mass binaries'' for systems with binary masses of about 2.7~$M_{\odot}$ to distinguish them from ``high-mass binaries'' with binary masses closer to or above the threshold mass.

The paper is organized as follows: In Sect.~\ref{sec:sim} we briefly review the simulations investigated in this study. Sect.~\ref{sec:idea} outlines the main idea. The method and its results are described in Sect.~\ref{sec:extra}. We close with a summary and conclusions.

\section{EoS and Simulations}\label{sec:sim}
In this study we further analyze the neutron-star merger simulations which have been presented in  \cite{2013PhRvL.111m1101B} and previous publications. The numerical calculations are performed with a relativistic smoothed particle hydrodynamics code, which incorporates the conformal flatness condition to solve for the spacetime metric~\cite{1980grg..conf...23I,1996PhRvD..54.1317W}. More details on the physical model and on the numerical implementation can be found in~\cite{2002PhRvD..65j3005O,2007A&A...467..395O,2010PhRvD..82h4043B,2012PhRvD..86f3001B}. Information on the general dynamics of the models, the convergence properties, and a comparison to fully general-relativistic studies are provided in~\cite{2012PhRvL.108a1101B,2012PhRvD..86f3001B,2013PhRvL.111m1101B} (see also~\cite{2013PhRvD..88d4026H,2014arXiv1403.5672T} for a comparison of the dominant GW oscillation frequency). At present, we do not include the effects of neutrino cooling or magnetic fields.

We consider microphysical, temperature-dependent EoSs (see Table~\ref{tab1}, which also provides the references to the individual models), which is essential for an accurate description of the merger process and of the stability properties of the remnant. The mass-radius relations resulting from these high-density models cover essentially the full range of possible stellar parameters, with cold, nonrotating $1.35~M_{\odot}$ models having radii in the range of 11.92~km to 14.74~km (see Fig.~\ref{fig:MR} for the mass-radius relations). The maximum masses of nonrotating NSs described by these EoSs vary between 1.94~$M_{\odot}$ and 2.79~$M_{\odot}$ (see Table~\ref{tab1}). Except for the IUF EoS, all high-density models are compatible with the current lower limit on the maximum mass of NSs given by~\cite{2010Natur.467.1081D,Antoniadis26042013}. Note that in contrast to our previous work, we do not include here the Shen~\cite{1998NuPhA.637..435S} and the GS1~\cite{2011PhRvC..83c5802S} EoSs because we use the TM1~\cite{1994NuPhA.579..557S,2012ApJ...748...70H} and the NL3~\cite{1997PhRvC..55..540L,2010NuPhA.837..210H} EoSs, which result in very similar mass-radius relationships.

For every EoS we perform simulations with systematically varied total binary mass $M_{\mathrm{tot}}$, focussing mostly on equal-mass systems. The GW emission is analyzed and the dominant postmerger GW frequency, which occurs as a pronounced peak in the GW spectrum, is extracted. For a consistent comparison between the different models, we consider a fixed duration of 10~ms after merger and apply a Hann window to the GW strain~\footnote{Certain models close to the instability limit have a lifetime shorter than 10~ms and thus the GW spectrum is only computed until collapse takes place.}. For some models, applying the Hann window causes a slight shift of the peak frequencies compared to previous publications~\cite{2012PhRvL.108a1101B,2012PhRvD..86f3001B,2013PhRvL.111m1101B}.

\section{MAIN idea}\label{sec:idea}
\begin{figure}
\includegraphics[width=8.9cm]{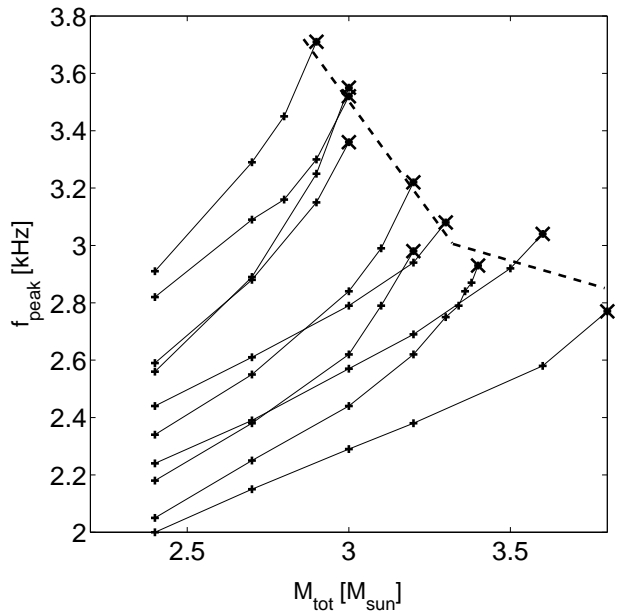}
\caption{\label{fig:f2mtot}Dominant GW frequency $f_{\mathrm{peak}}$ of the postmerger phase as a function of the total binary mass $M_{\mathrm{tot}}$ for all EoSs. Different EoSs are distinguished by different solid lines. The highest frequency $f_{\mathrm{peak}}^{\mathrm{thres}}$ for a given EoS is highlighted by a thick cross. The dashed line approximates the dependence of $f_{\mathrm{peak}}^{\mathrm{thres}}$ on the maximum binary mass $M_{\mathrm{thres}}$ which still produces an (at least transiently) stable merger remnant.}
\end{figure}
The idea underlying this study becomes clear from Fig.~\ref{fig:f2mtot}. It shows the dominant GW frequency as a function of the total binary mass for different EoSs considering equal-mass systems. Every EoS corresponds to one solid line, whose end point marks the most massive binary configuration ($M_{\mathrm{thres}}$) which leads to a differentially rotating NS merger remnant. (More massive binary systems collapse promptly to a BH on a dynamical time scale\footnote{The prompt collapse is identified by a continuous decrease of the minimum lapse function after the onset of the merging. The formation of a (possibly short-lived) NS merger remnant (``delayed collapse'') is distinguished by an increase of the minimum lapse after the initial drop and a (possibly small) number of subsequent oscillations in the minimum lapse function.}.) For a given EoS the dominant GW frequency of the postmerger phase increases with the total binary mass until it reaches the end point at a binary mass $M_{\mathrm{thres}}$ (the \textit{threshold mass} to BH collapse).

It was shown in~\cite{2013PhRvL.111m1101B} that there exists a relation between the binary mass $M_{\mathrm{thres}}$ (resulting in the most massive NS remnant) and the corresponding GW oscillation frequency $f_{\mathrm{peak}}^{\mathrm{thres}}$. (The crosses in Fig.~\ref{fig:f2mtot} are identical to the right panel of Fig.~3 in~\cite{2013PhRvL.111m1101B}). For all EoSs the data points $(M_{\mathrm{thres}}, f_{\mathrm{peak}}^{\mathrm{thres}})$ form a ``stability line'' (thick dashed line in Fig.~\ref{fig:f2mtot}) beyond which binary mergers lead to the direct formation of a BH. Our definition of the threshold mass $M_{\mathrm{thres}}$ is motivated by the observation that simulations with $M_{\mathrm{thres}}$ yield (at least transiently) stable remnants, whereas simulations with  $M_{\mathrm{thres}}+0.1M_\odot$  result in  prompt BH formation. (Note that in~\cite{2013PhRvL.111m1101B} an intermediate value of ${M_\mathrm{thres}}+0.05~M_{\odot}$ was denoted as $M_{\mathrm{thres}}$, which \textit{only reflects the 
uncertainty in determining } $M_{\mathrm{thres}}$ with our current set of simulated binary masses.)

\begin{figure}
\includegraphics[width=8.9cm]{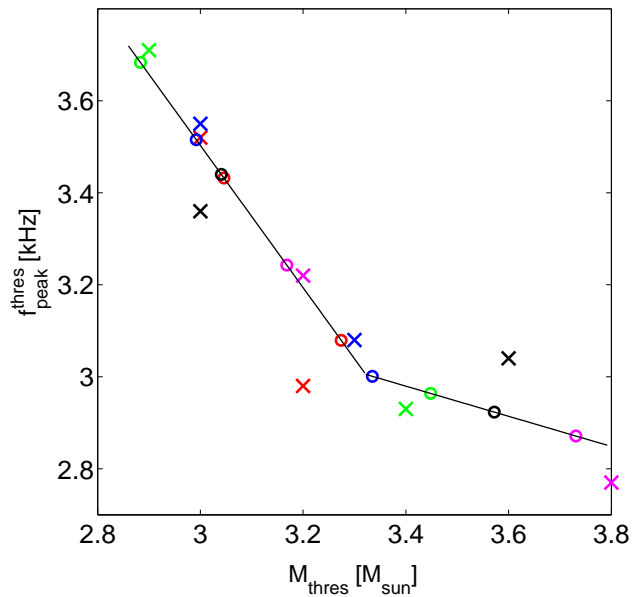}
\caption{\label{fig:fstabmstab}Dominant GW frequency $f_{\mathrm{peak}}^{\mathrm{thres}}$ of the most massive NS merger remnant as a function of the corresponding total binary mass $M_{\mathrm{thres}}$ for different EoSs (crosses). Circles of the same color denote the estimated values for $f_{\mathrm{peak}}^{\mathrm{thres}}$ and $M_{\mathrm{thres}}$ extrapolated entirely from GW information from low-mass NS binary mergers using Eqs.~\eqref{eq:extra} and~\eqref{eq:shift}. The solid line represents the fit to the stability limit (Eq.~\eqref{eq:stab}).}
\end{figure}

It has also been pointed out in~\cite{2013PhRvL.111m1101B} that the determination of $M_{\mathrm{thres}}$ and $f_{\mathrm{peak}}^{\mathrm{thres}}$ may yield important insights into the maximum mass of nonrotating NSs and on the radius of the maximum-mass configuration. As argued in the introduction, $M_{\mathrm{thres}}$ might be difficult to determine directly, because the merger of binary systems with masses near $M_{\mathrm{thres}}$ (which would be suitable for directly probing the approach to collapse), is expected to be less frequent, according to population synthesis studies and observations (e.g.~\cite{2012ARNPS..62..485L,2012ApJ...759...52D}). Moreover, several detections with different binary masses above and below the threshold would be required to deduce $M_{\mathrm{thres}}$ with a certain precision.

It is evident from Fig.~\ref{fig:f2mtot} that (at least) two measurements of $f_{\mathrm{peak}}$ at slightly different masses yield the slope $\frac{d f_{\mathrm{peak}}(M_{\mathrm{tot}})}{d M_{\mathrm{tot}}}$ and can be used for an extrapolation along the corresponding solid line. For a given EoS the extrapolation yields the intersection with the stability line, i.e. the line formed by the $M_{\mathrm{thres}}$ points for different EoSs (dashed line). In particular, to determine the slope $\frac{d f_{\mathrm{peak}}(M_{\mathrm{tot}})}{d M_{\mathrm{tot}}}$, detections in the mostly likely range of binary parameters with $M_{\mathrm{tot}}\sim 2.7~M_{\odot}$ can be employed.
\begin{figure}
\includegraphics[width=8.9cm]{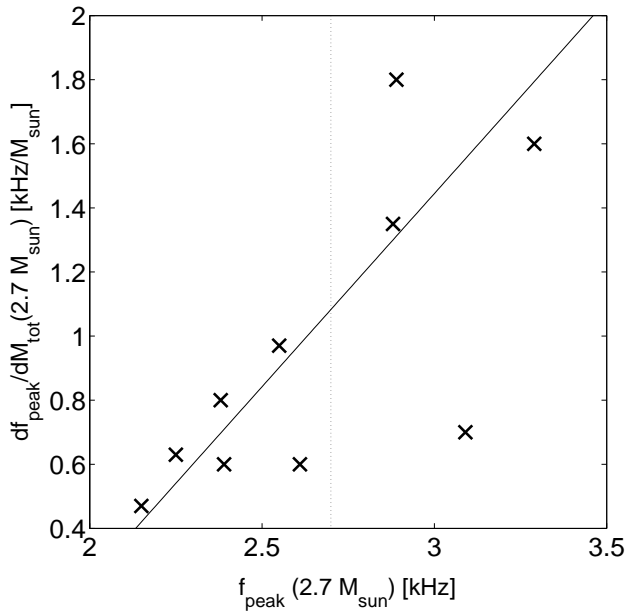}
\caption{\label{fig:slope}Derivative of $f_{\mathrm{peak}}(M_{\mathrm{tot}})$ with respect to $M_{\mathrm{tot}}$ at $M_{\mathrm{tot}}=2.7~M_{\odot}$ as a function of $f_{\mathrm{peak}}(2.7~M_{\odot})$ for different EoSs. The thin solid line separates models with a normal/steep relative slope (above) from models with flat relative slopes (below). The vertical dotted line distinguishes low frequencies and high frequencies (see main text).}
\end{figure}
In Fig.~\ref{fig:f2mtot} we notice that for all sequences of different EoSs the slope (of the solid lines) becomes steeper towards the (dashed) stability line at $M_{\mathrm{thres}}$. Hence, a \textit{linear extrapolation} in general will tend to overestimate $M_{\mathrm{thres}}$ and underestimate the corresponding $f_{\mathrm{peak}}^{\mathrm{thres}}$.

The increasing slope with the binary mass can be understood because the dominant GW emission of the postmerger phase is produced by the fundamental quadrupolar ($m=2$) fluid mode~\cite{2011MNRAS.418..427S}, whose frequency scales approximately with the mean density, i.e. $\sqrt{M_{\mathrm{remnant}}/R_{\mathrm{remnant}}^3}$~\cite{1998MNRAS.299.1059A,2012PhRvL.108a1101B}. For a given EoS, radii of massive NSs decrease with mass, which explains the steeper increase of $f_{\mathrm{peak}}$ at higher $M_{\mathrm{tot}}$.

The main idea of this work is to introduce an \textit{extrapolation procedure}, which employs GW detections of binaries at masses of about $2.7M_{\odot}$, in order to estimate the properties of mergers at higher masses. Throughout this paper, crosses mark data which have been obtained by numerical calculations and are considered to be the ``true'' (actual) values for a given EoS. Circles are used whenever a quantity is estimated by means of the extrapolation method, i.e. when only information from GW measurements of mergers with total binary masses of about~$2.7M_{\odot}$ are used to estimate its value.

\section{Extrapolation procedure}\label{sec:extra}
\subsection{Predicting the threshold mass and the maximum mass}
In this paper we explore what can be inferred from (at least) two measurements of the dominant postmerger GW frequency at two relatively low, slightly different binary masses in the range of about 2.7~$M_{\odot}$. We thus work under the condition that the frequency and the slope $\frac{d f_{\mathrm{peak}}}{d M_{\mathrm{tot}}}$ at $M_{\mathrm{tot}}=2.7~M_{\odot}$ have been determined observationally (see~\cite{2013arXiv1312.1862M,Clark2014}).

The stability line (dashed line in Fig.~\ref{fig:f2mtot}) can be analytically approximated by a simple broken straight line, which is obtained by a fit to the data points with $f_{\mathrm{peak}}^{\mathrm{thres}}>3.1$~kHz and another fit for the data points located at lower frequencies. The function $f_{\mathrm{peak}}^{\mathrm{thres}}(M_{\mathrm{thres}})$ is then obtained as:
\begin{equation} \label{eq:stab}
f_{\mathrm{peak}}^{\mathrm{thres}}(M_{\mathrm{thres}})=\begin{cases}
-0.328\cdot M_{\mathrm{thres}} +4.093, \\
 \ \ \ \ \ \ \ \ \ \ \ \ \ \ \ \ \ \ \ \ \ \ \ \ \ M_{\mathrm{thres}}<3.32~M_{\odot}&\\
-1.546\cdot M_{\mathrm{thres}} +8.140, \\  \ \ \ \ \ \ \ \ \ \ \ \ \ \ \ \ \ \ \ \ \ \ \ \ \    M_{\mathrm{thres}}>3.32~M_{\odot}&\\
\end{cases}
\end{equation}
with $M_{\mathrm{thres}}$ in solar masses and $f_{\mathrm{peak}}^{\mathrm{thres}}$ measured in kHz.  Note that the intersection of the two line segments (as defined above) occurs below 3.1~kHz. After obtaining the two line segments, the individual $(f_{\mathrm{peak}},M_{\mathrm{tot}})$ data points at $M_{\mathrm{tot}}\approx 2.7~M_{\odot}$ are extrapolated linearly towards the stability line.

A simple linear extrapolation onto the curve given by Eq.~\eqref{eq:stab} certainly underestimates $M_{\mathrm{thres}}$, because the slope becomes steeper towards the stability line (see Fig.~\ref{fig:f2mtot}). We obtain a better estimate, by accounting for this bending by slightly shifting upwards the extrapolating line. Specifically, we find an upwards shift of 0.1~kHz to satisfactorily counteract the bending of the lines.

The estimates for $M_{\mathrm{thres}}$ can be further refined for models which are relatively far away from the intersection with the stability line (i.e. models with low peak frequencies $f_{\mathrm{peak}}(2.7~M_{\odot})$). At a fixed binary mass of 2.7~$M_{\odot}$ in general the gradient is larger for higher $f_{\mathrm{peak}}$. This can be seen in Fig.~\ref{fig:slope}, which shows for all EoSs the slope as a function of the peak frequency at $M_{\mathrm{tot}}=2.7~M_{\odot}$. However, at lower frequencies one observes in Fig.~\ref{fig:f2mtot} that EoSs which have a relatively flat slope with respect to their peak frequency, tend to extend slightly to the right of the stability line. To accommodate this behavior, we found empirically that it is useful to \textit{choose a shift which depends on the relative slope}. 

Thus, our final recipe is to apply for models with $f_{\mathrm{peak}}(2.7~M_{\odot})<2.7$~kHz (on the left of the vertical dotted line in Fig.~\ref{fig:slope}) a shift of 0.01~kHz, if they have a \textit{relatively small slope}, while we use a shift of 0.24~kHz for a \textit{steeper/normal increase} of $f_{\mathrm{peak}}(M_{\mathrm{tot}})$. The shift for the steep/normal slope for models with $f_{\mathrm{peak}}(2.7~M_{\odot})<2.7$ is higher than our default choice, because the extrapolation is performed over a larger interval and thus has to compensate an overall stronger bending. The models which show a relatively small slope can be identified in Fig.~\ref{fig:slope} as the points located below the approximately straight line formed by the majority of EoSs. Quantitatively, we distinguish whether a data point below 2.7~kHz is above or below the line $1.21\cdot f_{\mathrm{peak}} -2.17$ (indicated in Fig.~\ref{fig:slope}). 

Using the definitions $f_{2.7}=f_{\mathrm{peak}}(2.7~M_{\odot})$ and ${f}'_{2.7}=\frac{d f_{\mathrm{peak}}}{d M_{\mathrm{tot}}}(2.7~M_{\odot})$ the extrapolating curve thus reads
\begin{equation}\label{eq:extra}
f_{\mathrm{peak}}(M_{\mathrm{tot}}) = f_{2.7} + {f}'_{2.7} (M_{\mathrm{tot}}-2.7~M_{\odot}) + \sigma
\end{equation}
with the shift $\sigma$ given by
\begin{equation}\label{eq:shift}
\sigma=\begin{cases}
0.1,~ f_{2.7}>2.7~\mathrm{kHz} \\
0.01,~ f_{2.7}<2.7~\mathrm{kHz~ and~ } {f}'_{2.7}<1.21 f_{2.7}-2.17 \\
0.24,~ f_{2.7}<2.7~\mathrm{kHz~ and~ } {f}'_{2.7}\geq 1.21 f_{2.7}-2.17. \\
\end{cases}
\end{equation}

It is important to note that the results are rather insensitive to the exact details of this extrapolation method or to the precise fit to approximate the stability line. Such details influence the final results only on the level of a few percent. Further improvements may certainly be possible by more 
sophisticated procedures, but our study is meant as a proof of principle and refinements only make sense when more EoSs are considered (even though our sample already extends over a rather wide range of candidates EoSs). 

The results of the extrapolation procedure described in detail above, are summarized in Table~\ref{tab1}. Figure~\ref{fig:fstabmstab} visualizes the estimated (circles) and the actual values (crosses). \textit{We find that the extrapolated values for  $M_{\mathrm{thres}}$ match the actual $M_{\mathrm{thres}}$ with an accuracy of 2\%\ or better}. Hence, the threshold mass can be estimated with high accuracy from extrapolations of GW detections of lower-mass binaries (with total mass in the range of about 2.7~$M_{\odot}$).

The above estimate of the threshold mass   can now be converted to a determination of the maximum mass of cold, nonrotating NSs by an inversion of the empirical fit
\begin{equation}\label{eq:mthr}
M_{\mathrm{thres}}=\left( -3.606\cdot\frac{G M_{\mathrm{max}}}{c^2 R_{1.6}}+2.380\right) M_{\mathrm{max}}-0.05~M_{\odot},
\end{equation}
which was recently found in~\cite{2013PhRvL.111m1101B} to describe the relation between these two masses, independently of the EoS, with an accuracy of at least 0.1~$M_{\odot}$. (Masses in Eq.~\eqref{eq:mthr} are in solar masses and radii are in km). $R_{1.6}$ denotes the radius of a cold, nonrotating NS with 1.6~$M_{\odot}$.

\begin{figure}
\includegraphics[width=8.9cm]{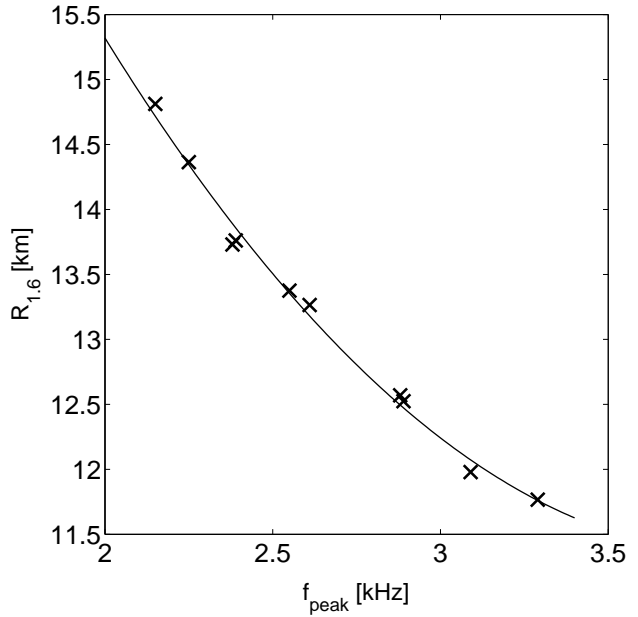}
\caption{\label{fig:fpeak}Relation between the radii $R_{1.6}$ of cold, nonrotating NSs with a gravitational mass of 1.6~$M_{\odot}$ and the dominant postmerger GW frequency $f_{\mathrm{peak}}$ for a total binary mass of  2.7~$M_{\odot}$ for different EoSs. The line displays a quadratic fit to the data.}
\end{figure}

In turn, the radius $R_{1.6}$ is given by the peak frequency measured at $M_{\mathrm{tot}}=2.7~M_{\odot}$~\cite{2012PhRvL.108a1101B,2012PhRvD..86f3001B}, which exhibits a tight correlation with $R_{1.6}$, as shown in Fig.~\ref{fig:fpeak}.~\footnote{Based on a smaller set of investigated EoSs, in Ref.~\cite{2013PhRvD..88d4026H} it was suggested that a fiducial NS mass of 1.8~$M_{\odot}$ yields the smallest scatter in the relation between the peak frequency and the radii of nonrotating NS. Which exact fiducial mass leads to the smallest scatter certainly depends somewhat on the exact set of EoSs under consideration given that the radii of NSs with masses of 1.6~$M_{\odot}$ and 1.8~$M_{\odot}$ are very similar.}
\begin{figure}
\includegraphics[width=8.9cm]{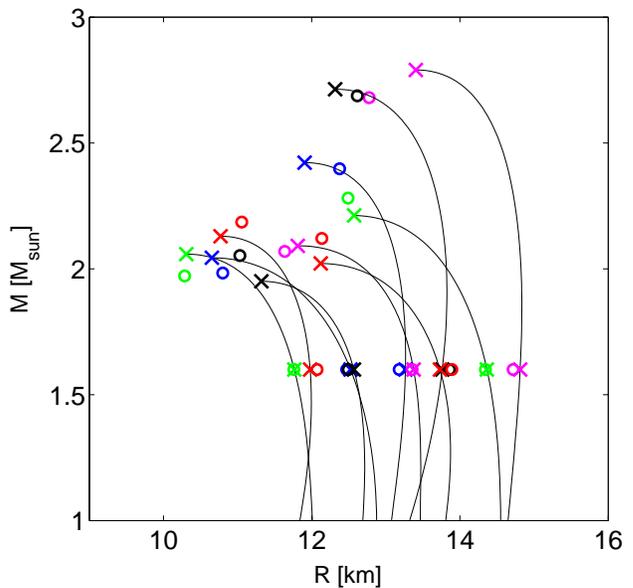}
\caption{\label{fig:MRpoints}Mass-radius relations for the EoSs considered in this study with the circumferential radius $R$ and the gravitational mass $M$. Crosses at higher masses mark maximum-masses of nonrotating NSs. Circles of the same color indicate the values estimated by the extrapolation of GW data of low-mass NS mergers and use of Eq.~\eqref{eq:mthr} to obtain $M_{\mathrm{max}}$ from the $M_{\mathrm{thres}}$ estimate. The crosses at 1.6~$M_{\odot}$ mark the radii of 1.6~$M_{\odot}$ NSs. Circles at 1.6~$M_{\odot}$ indicate the radius estimate via the dominant postmerger GW frequency of a 2.7~$M_{\odot}$ NS merger.}
\end{figure}
We implement the empirical relation seen in Fig.~\ref{fig:fpeak} by a quadratic least-squares fit, which yields
\begin{equation} \label{eq:fpeak}
R_{1.6} = 1.099\cdot f_{\mathrm{peak}}^2 - 8.574\cdot f_{\mathrm{peak}} +28.07.
\end{equation}
This fit slightly deviates from the expression given in~\cite{2012PhRvL.108a1101B} because here we consider a different set of representative EoSs, taking into account only the fully temperature-dependent EoSs.
\begin{figure*}

\includegraphics[width=15cm]{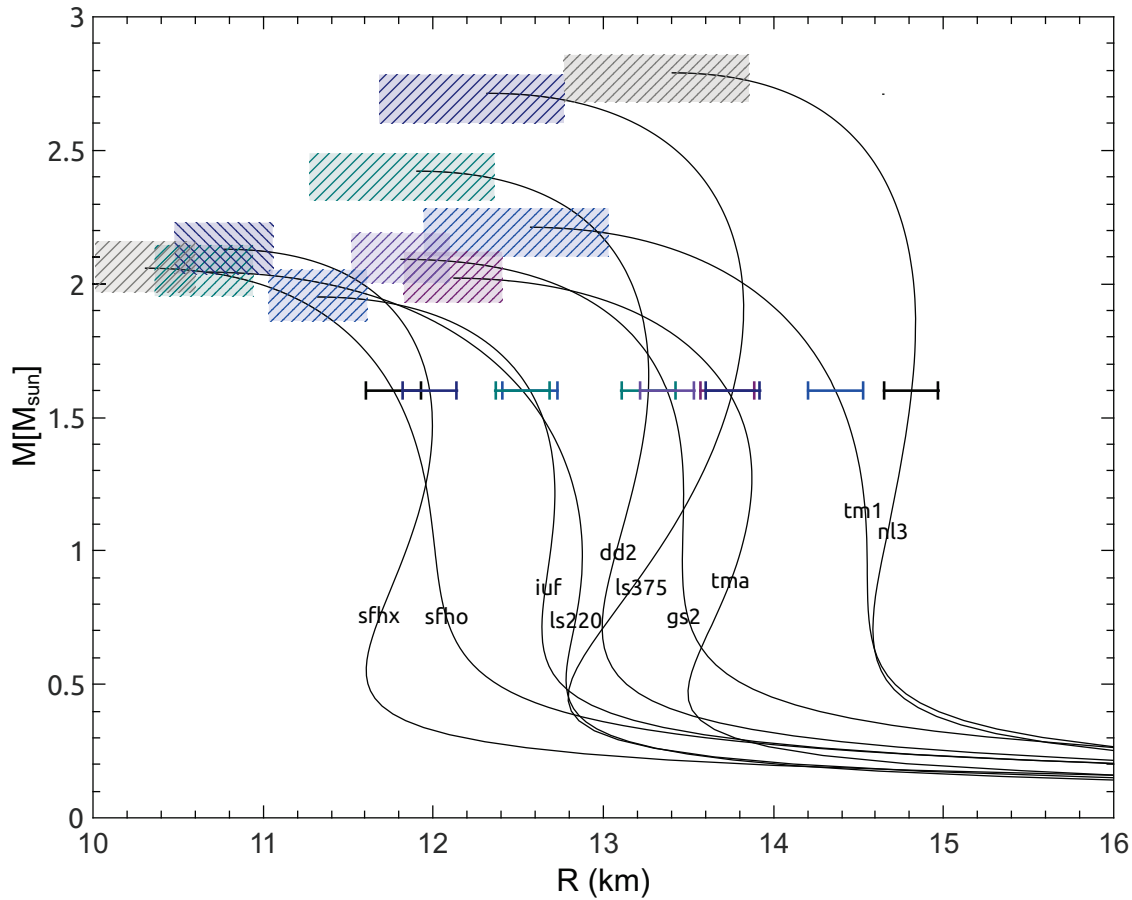}
\caption{\label{fig:MR} Mass-radius relations for the EoSs considered in this study with the radius $R$ and the gravitational mass $M$. Boxes illustrate the maximum deviation of the estimated properties of the maximum-mass configuration, which are inferred from GW detections of low-mass binary NS mergers. The size of the boxes is chosen to be {\it the largest deviation found in the sample of EoSs with low maximum masses ($M_{\mathrm{max}}<2.2~M_{\odot}$) and the sample of EoSs with high maximum masses ($M_{\mathrm{max}}>2.2~M_{\odot}$)}. Bars at 1.6~$M_{\odot}$ indicate the maximum deviation of the estimated radius inferred from a single GW detection of a low-mass binary NS merger. The size of the bars is chosen to be the {\it largest deviation from the actual value found in the whole sample of EoSs}.}
\end{figure*}
\begin{figure}
\includegraphics[width=8.7cm]{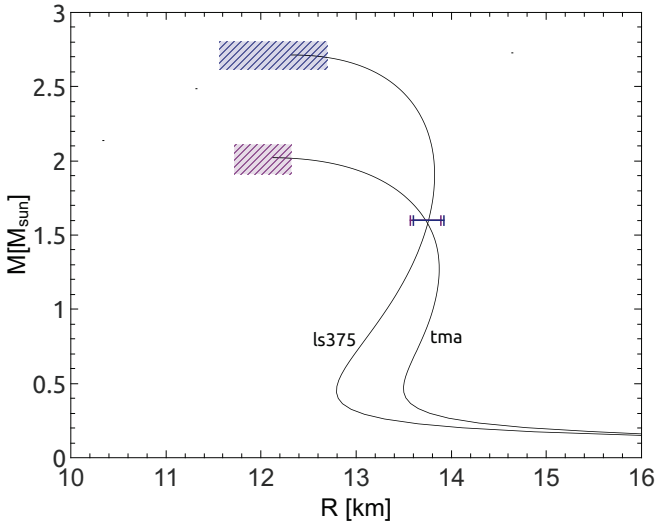}
\caption{\label{fig:MR2}Same as Fig.~\ref{fig:MR} for two EoSs with similar stellar properties in the intermediate mass range around 1.6~$M_{\odot}$ where the two mass-radius relations cross. Using the extrapolation procedure described in the main text (Sect.~\ref{sec:extra}) the two EoSs can clearly be distinguished.}
\end{figure}

Finally, using Eq.~\eqref{eq:fpeak} the inversion of Eq.~\eqref{eq:mthr} recovers the maximum mass with a maximum deviation of~0.1~$M_{\odot}$ from the actual value (see Table~\ref{tab1}). The estimated and the actual values of the maximum-mass configuration are shown in Fig.~\ref{fig:MRpoints}. The estimates for the corresponding radii of the maximum-mass configuration are presented below. We visualize the deviations of the estimated values from the actual values in Fig.~\ref{fig:MR} by drawing boxes around the actual maximum-mass configurations such that they include the estimated values. \textit{The width and height of the boxes are defined by the overall maximum positive and negative deviation of estimated masses and radii from their actual values, within the sample of EoSs we consider}. Here, we distinguish EoSs which terminate on the lower segment of the stability line in Fig.~\ref{fig:f2mtot} from EoSs that terminate on the upper line segment. The latter models exhibit smaller deviations in their estimated radii.

Note  that for many individual EoSs the deviations of the estimated mass and radius of the maxium-mass model are significantly smaller than indicated by the boxes (see Fig.~\ref{fig:MRpoints}). Also, note that for the models of the upper branch of Eq.~\eqref{eq:stab}
the largest deviations are found for the IUF and the TMA EoSs, which are already excluded or marginally compatible with pulsar observations~\cite{2010Natur.467.1081D,Antoniadis26042013}. The remaining EoSs of the upper branch appear to have smaller deviations, which can be understood because the stability line is steeper and closer to the measured data at $M_{\mathrm{tot}}=2.7~M_{\odot}$.

Moreover, in Fig. \ref{fig:MR} we use horizontal error bars to indicate the deviations of the $R_{1.6}$ estimate via $f_{\mathrm{peak}}(M_{\mathrm{tot}}=2.7~M_{\odot})$ when employing Eq.~\eqref{eq:fpeak}. A powerful feature of our method is that it allows to distinguish (by estimating the maximum mass model)\ two EoSs that cross at around $1.6~M_{\odot}$. This is demonstrated clearly in Fig. \ref{fig:MR2},  where the two EoSs LS375 and TMA have very similar radii of about 13.8~km at 1.6~$M_{\odot}$ (and thus an individual radius estimate, based on only Eq.~(\ref{eq:fpeak}) would be degenerate with respect to the underlying EoS). In contrast, having also the estimate on the mass and radius of the maximum mass model (based on the novel extrapolation procedure described above) clearly distinguishes the two EoSs. More examples of this type can be identified in Fig.~\ref{fig:MR}.

Note that the extrapolation proposed here is also useful to identify \textit{lower and upper limits} on the maximum mass of cold, nonrotating NS. The data points forming the stability line in Fig.~\ref{fig:f2mtot} can be embraced by displacing Eq.~\eqref{eq:stab} downwards by 0.2~kHz, to obtain  a lower limit and by adding 0.2~kHz to obtain an upper limit consistent with our current sample of models.

\begin{table*}
\caption{\label{tab1} Equation of state models with references and resulting stellar properties. $M_{\mathrm{max}}$ denotes the maximum mass of nonrotating NSs with the cirumferential radius $R_{\mathrm{max}}$ corresponding this maximum-mass configuration. $e_{\mathrm{max}}$ and $\rho_{\mathrm{max}}$ are the central energy density and the central rest-mass density of the maximum-mass configuration. $R_{1.6}$ refers to the circumferential radius of a nonrotating 1.6~$M_{\odot}$ NS. $M_{\mathrm{thres}}$ is the highest total binary mass which leads to differentially rotating NS merger remnant for the given EoS. The dominant GW frequency of this postmerger remnant is $f_{\mathrm{peak}}^{\mathrm{thres}}$. Hatted quantities are the estimates for these merger properties and stellar parameters based on the extrapolation procedure described in the main text (Sect.~\ref{sec:extra}).}
 \begin{ruledtabular}
 \begin{tabular}{|l|l|l|l|l| l|l|l|l|l| l|l|l|l|l|}
\hline    
& $M_{\mathrm{max}}$ & $\hat{M}_{\mathrm{max}}$ & $R_{1.6}$ & $\hat{R}_{1.6}$ & $M_{\mathrm{thres}}$ & $\hat{M}_{\mathrm{thres}}$ & $f_{\mathrm{peak}}^{\mathrm{thres}}$ & $\hat{f}_{\mathrm{peak}}^{\mathrm{thres}}$ & $R_{\mathrm{max}}$ & $\hat{R}_{\mathrm{max}}$ & $e_{\mathrm{c,max}}$ & $\hat{e}_{\mathrm{c,max}}$ & $\rho_{\mathrm{c,max}}$ & $\hat{\rho}_{\mathrm{c,max}}$ \\
EoS   & $(M_{\odot})$ & $(M_{\odot})$ & (km) & (km) &  $(M_{\odot})$ & $(M_{\odot})$ & (kHz) & (kHz) & (km) & (km) & $(\mathrm{g/cm^3})$ & $(\mathrm{g/cm^3})$ & $(\mathrm{g/cm^3})$ & $(\mathrm{g/cm^3})$\\ \hline
%Eos                                                    mmax              R16                 Mstab            Fstab             Rmax                emax                            rhomax
NL3~\cite{1997PhRvC..55..540L,2010NuPhA.837..210H}   &  2.79  &  2.68  &  14.81  &   14.72 &  3.8  &   3.73 &  2.77  &  2.87 &  13.40  &   12.78  &  1.52$\times 10^{15}$  & 1.68 $\times 10^{15}$ &  1.09$\times 10^{15}$  & 1.25$\times 10^{15}$  \\
LS375~\cite{1991NuPhA.535..331L}                     &  2.71  &  2.69  &  13.76  &   13.86 &  3.6  &   3.57 &  3.04  &  2.93 &  12.32  &   12.62  &  1.78$\times 10^{15}$  & 1.74 $\times 10^{15}$ &  1.25$\times 10^{15}$  & 1.29$\times 10^{15}$  \\
DD2~\cite{2010PhRvC..81a5803T,2010NuPhA.837..210H}   &  2.42  &  2.40  &  13.26  &   13.18 &  3.3  &   3.33 &  3.08  &  3.00 &  11.90  &   12.38  &  1.95$\times 10^{15}$  & 1.83 $\times 10^{15}$ &  1.41$\times 10^{15}$  & 1.35$\times 10^{15}$  \\
TM1~\cite{1994NuPhA.579..557S,2012ApJ...748...70H}   &  2.21  &  2.28  &  14.36  &   14.34 &  3.4  &   3.45 &  2.93  &  2.96 &  12.57  &   12.49  &  1.80$\times 10^{15}$  & 1.79 $\times 10^{15}$ &  1.36$\times 10^{15}$  & 1.32$\times 10^{15}$  \\
SFHX~\cite{2012arXiv1207.2184S}                      &  2.13  &  2.19  &  11.98  &   12.07 &  3.0  &   3.05 &  3.52  &  3.43 &  10.77  &   11.06  &  2.39$\times 10^{15}$  & 2.33 $\times 10^{15}$ &  1.74$\times 10^{15}$  & 1.71$\times 10^{15}$  \\
GS2~\cite{2011arXiv1103.5174S}                       &  2.09  &  2.07  &  13.38  &   13.35 &  3.2  &   3.17 &  3.22  &  3.24 &  11.81  &   11.64  &  2.05$\times 10^{15}$  & 2.11 $\times 10^{15}$ &  1.56$\times 10^{15}$  & 1.55$\times 10^{15}$  \\
SFHO~\cite{2012arXiv1207.2184S}                      &  2.06  &  1.97  &  11.77  &   11.76 &  2.9  &   2.88 &  3.71  &  3.68 &  10.31  &   10.29  &  2.67$\times 10^{15}$  & 2.63 $\times 10^{15}$ &  1.91$\times 10^{15}$  & 1.92$\times 10^{15}$  \\
LS220~\cite{1991NuPhA.535..331L}                     &  2.04  &  1.98  &  12.52  &   12.47 &  3.0  &   2.99 &  3.55  &  3.52 &  10.65  &   10.80  &  2.55$\times 10^{15}$  & 2.43 $\times 10^{15}$ &  1.86$\times 10^{15}$  & 1.78$\times 10^{15}$  \\
TMA~\cite{1995NuPhA.588..357T,2012ApJ...748...70H}   &  2.02  &  2.12  &  13.73  &   13.89 &  3.2  &   3.27 &  2.98  &  3.08 &  12.12  &   12.14  &  1.92$\times 10^{15}$  & 1.92 $\times 10^{15}$ &  1.48$\times 10^{15}$  & 1.42$\times 10^{15}$  \\
IUF~\cite{2010PhRvC..82e5803F,2010NuPhA.837..210H}   &  1.95  &  2.05  &  12.57  &   12.50 &  3.0  &   3.04 &  3.36  &  3.44 &  11.32  &   11.03  &  2.19$\times 10^{15}$  & 2.34 $\times 10^{15}$ &  1.67$\times 10^{15}$  & 1.72$\times 10^{15}$  \\

\end{tabular}
 \end{ruledtabular} 
\end{table*}

\subsection{Estimating the radius of the maximum-mass configuration}
\begin{figure}
\includegraphics[width=8.9cm]{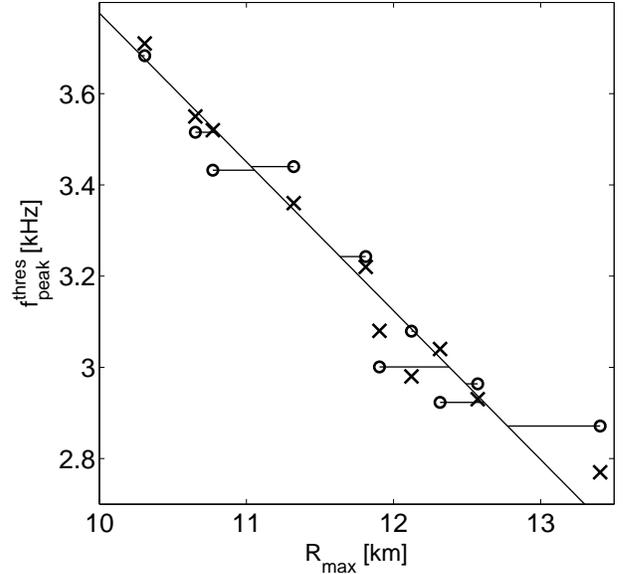}
\caption{\label{fig:rmax}Dominant GW frequency $f_{\mathrm{peak}}^{\mathrm{thres}}$ of the most massive NS merger remnant as a function of the radius $R_{\mathrm{max}}$ of the maximum-mass configuration of cold, nonrotating NSs for different EoSs (crosses). The diagonal solid line is a fit to $R_{\mathrm{max}}(f_{\mathrm{peak}}^{\mathrm{thres}})$. Circles denote the estimated values for $f_{\mathrm{peak}}^{\mathrm{thres}}$,  estimated entirely from GW information from low-mass NS binary mergers. The estimated values for $R_{\mathrm{max}}$ can be inferred by projecting horizontally, i.e. following the short lines to the diagonal line representing the fit to the numerical data (crosses).}
\end{figure}
As mentioned already in the previous sections, $M_{\mathrm{thres}}$ or $f_{\mathrm{peak}}^{\mathrm{thres}}$ determine also other stellar properties of NSs~\cite{2013PhRvL.111m1101B} and we proceed by discussing further insights that can be obtained, by applying our extrapolation method of GW information obtained from low-mass binary NS mergers. The intersection of the curves in Fig.~\ref{fig:f2mtot} with the stability line also provides an estimate for the GW oscillation frequency at $M_{\mathrm{thres}}$. This peak frequency $f_{\mathrm{peak}}^{\mathrm{thres}}$ scales well with the radius $R_{\mathrm{max}}$ of the maximum-mass configuration of cold, nonrotating NSs (see left panel of Fig.~3 in~\cite{2013PhRvL.111m1101B} and Fig.~\ref{fig:rmax}). (The relation can be understood by noting that $f_{\mathrm{peak}}^{\mathrm{thres}}$ should scale approximately with $\sqrt{M_{\mathrm{thres}}/R_{\mathrm{max}}^3}$, where the variation in $R_{\mathrm{max}}^3$ dominates over the relatively small change in 
$M_{\mathrm{thres}}$.) In Fig.~\ref{fig:rmax} we display the extrapolated $f_{\mathrm{peak}}^{\mathrm{thres}}$ 
(circles) and the actual frequency obtained in the simulations (crosses) as a function of $R_{\mathrm{max}}$ for different EoSs. Using the linear fit to the simulation data
\begin{equation}\label{eq:rmax}
R_{\mathrm{max}}=-3.065\cdot f_{\mathrm{peak}}^{\mathrm{thres}} +21.57 ~(\pm 0.7),
\end{equation}
the extrapolated frequency determines the radius of the maximum-mass configuration with an accuracy of typically 4\%\ or better. Only for the NL3 EoS the estimated $R_{\mathrm{max}}$ deviates by 5\%. The somewhat larger difference is understandable, considering that for NL3 the extrapolation is performed over the largest distance between data measured at 2.7~$M_{\odot}$ and at the intersection at $M_{\mathrm{thres}}\approx 3.8~M_{\odot}$). The results of the extrapolation procedure are listed in Table~\ref{tab1}, together with the actual values of $R_{\mathrm{max}}$. The estimated and actual radii of the maximum-mass configuration are also shown in~Fig.~\ref{fig:MRpoints}. The shifts denoted in parentheses in Eq.~\eqref{eq:rmax} define curves which lead to \textit{upper and lower limits} for $R_{\mathrm{max}}$, when used in the extrapolation procedure.

\subsection{Estimating the maximum central density}
\begin{figure}
\includegraphics[width=8.9cm]{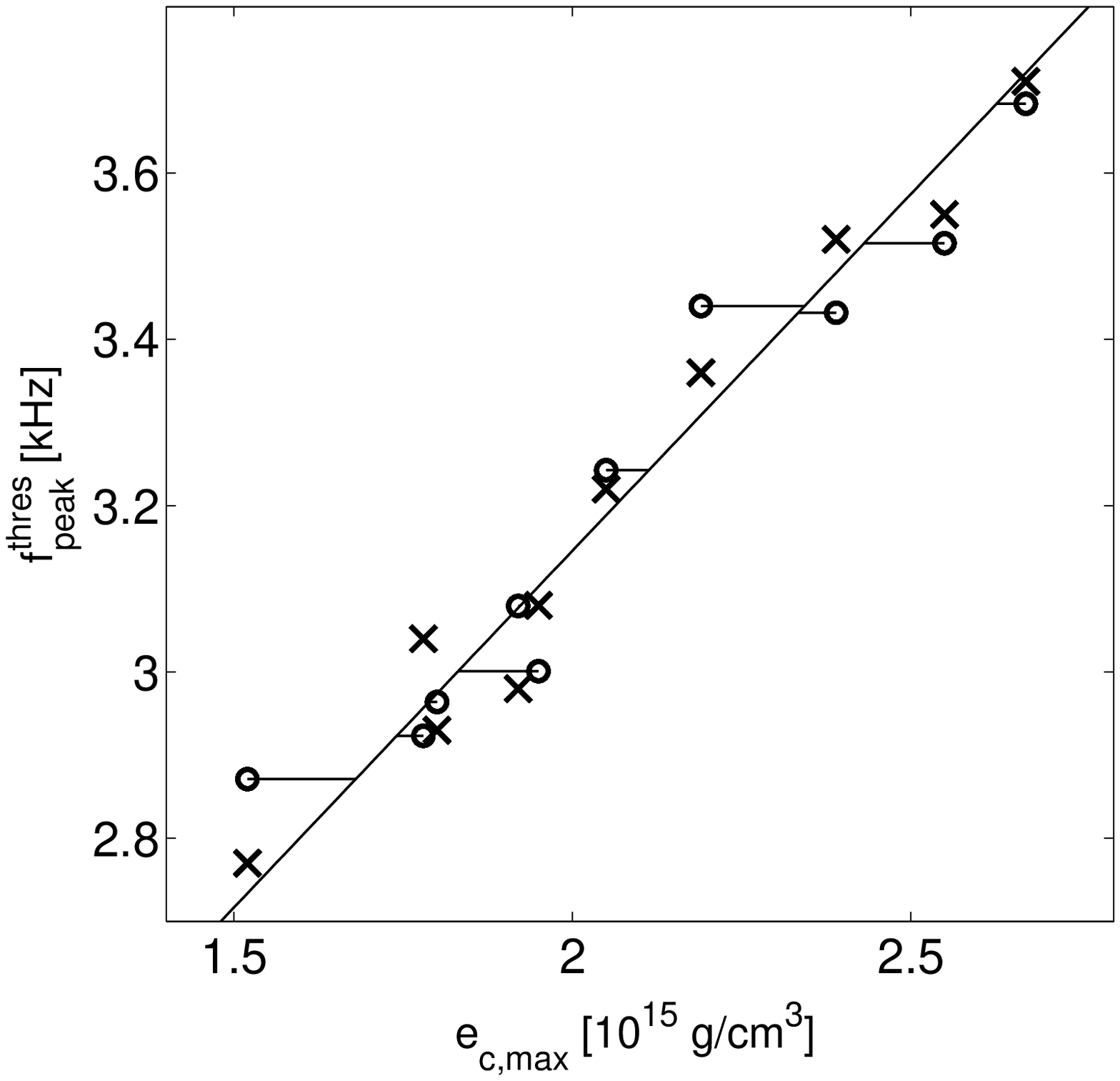}
\caption{\label{fig:emax}Dominant GW frequency $f_{\mathrm{peak}}^{\mathrm{thres}}$ of the most massive NS merger remnant as a function of the maximum central energy density $e_{\mathrm{c,max}}$ of the maximum-mass configuration of nonrotating NSs for different EoSs (crosses). The diagonal solid line is a fit to $e_{\mathrm{c,max}}(f_{\mathrm{peak}}^{\mathrm{thres}})$. Circles denote the estimated values for $f_{\mathrm{peak}}^{\mathrm{thres}}$,  extrapolated entirely from GW information from low-mass NS binary mergers. The estimated values for $e_{\mathrm{c,max}}$ can be inferred by projecting horizontally, i.e. following the short lines to the diagonal line representing the fit to the numerical data (crosses).}
\end{figure}
\begin{figure}
\includegraphics[width=8.9cm]{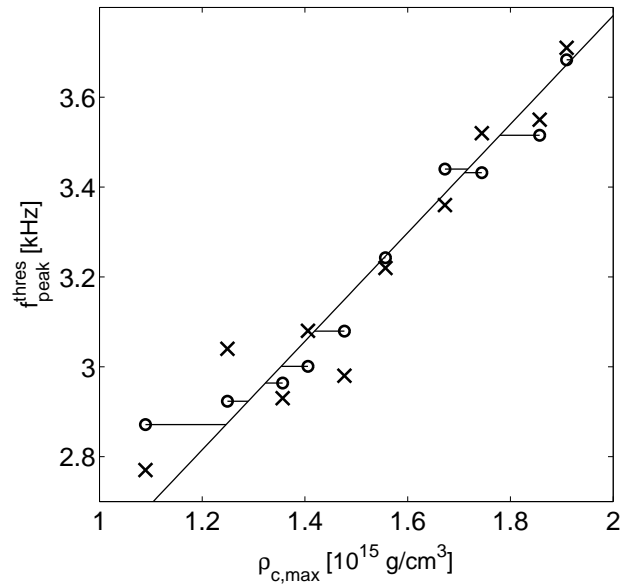}
\caption{\label{fig:rhomax}Dominant GW frequency $f_{\mathrm{peak}}^{\mathrm{thres}}$ of the most massive NS merger remnant as a function of the maximum central rest-mass density $\rho_{\mathrm{c,max}}$ of the maximum-mass configuration of nonrotating NSs for different EoSs (crosses). The diagonal solid line is a fit to $\rho_{\mathrm{c,max}}(f_{\mathrm{peak}}^{\mathrm{thres}})$. Circles denote the estimated values for $f_{\mathrm{peak}}^{\mathrm{thres}}$, extrapolated entirely from GW information from low-mass NS binary mergers. The estimated values for $\rho_{\mathrm{c,max}}$ can be inferred by projecting horizontally, i.e. following the short line to the diagonal line representing the fit to the numerical data (crosses).}
\end{figure}
For maximum-mass TOV solutions it is empirically known and intuitive that the stiffness of an EoS, quantified by the ratio $\langle e\rangle {}_{\mathrm{max}} /e_{\mathrm{c,max}}$ between the mean density and the central density, roughly scales linearly with the compactness $C_{\mathrm{max}}=\frac{G M_{\mathrm{max}}}{c^2 R_{\mathrm{max}}}$~\cite{1997ApJ...488..799K,2012ARNPS..62..485L} (see also Fig.~2 in~\cite{2013PhRvL.111m1101B}). Here, $e$ refers to the energy density, which, however, is related to the rest-mass density through the EoS and therefore, the following analysis yields analogous  results when applied to the rest-mass density (see Table~\ref{tab1}).

Adopting $\langle e\rangle {}_{\mathrm{max}}=\frac{3}{4\pi}\frac{M_{\mathrm{max}}}{R_{\mathrm{max}}^3}$ implies that the central density should scale roughly as $1/R_{\mathrm{max}}^2$. Consequently, it is possible to employ our extrapolation method to estimate the maximum central density of NSs and to establish lower and upper limits. The linear relation in Fig.~\ref{fig:rmax} suggests a relation between $e_{\mathrm{c,max}}$ and $f_{\mathrm{peak}}^{\mathrm{thres}}$, which is shown in Fig.~\ref{fig:emax}. In addition, Fig.~\ref{fig:emax} provides $f_{\mathrm{peak}}^{\mathrm{thres}}$, estimated with the extrapolation of GW data measured in low-mass NS binary mergers. Again, we employ a linear fit to the (actual) simulation data (crosses) and convert the extrapolated values for $f_{\mathrm{peak}}^{\mathrm{thres}}$ to an estimate for $e_{\mathrm{c,max}}$. The function fitting the data is given by
\begin{equation}\label{eq:fite}
e_{\mathrm{c,max}}=1.166\cdot f_{\mathrm{peak}}^{\mathrm{thres}} -1.668 ~(\pm 0.2).
\end{equation}
Here, $e_{\mathrm{c,max}}$ is given in $10^{15}\mathrm{g/cm^3}$, while frequencies are measured in kHz.

In Table~\ref{tab1}, the estimated central energy densities are compared with the actual ones. The estimated values and the actual values agree within 7\% (except for the NL3 EoS, which deviates by 11\%). By using fit formulae that embrace the numerical data (shifting the fit in Eq.~\eqref{eq:fite} by $\pm 0.2$~kHz) one can define upper and lower limits for the given set of EoSs.
\begin{figure}
\includegraphics[width=8.9cm]{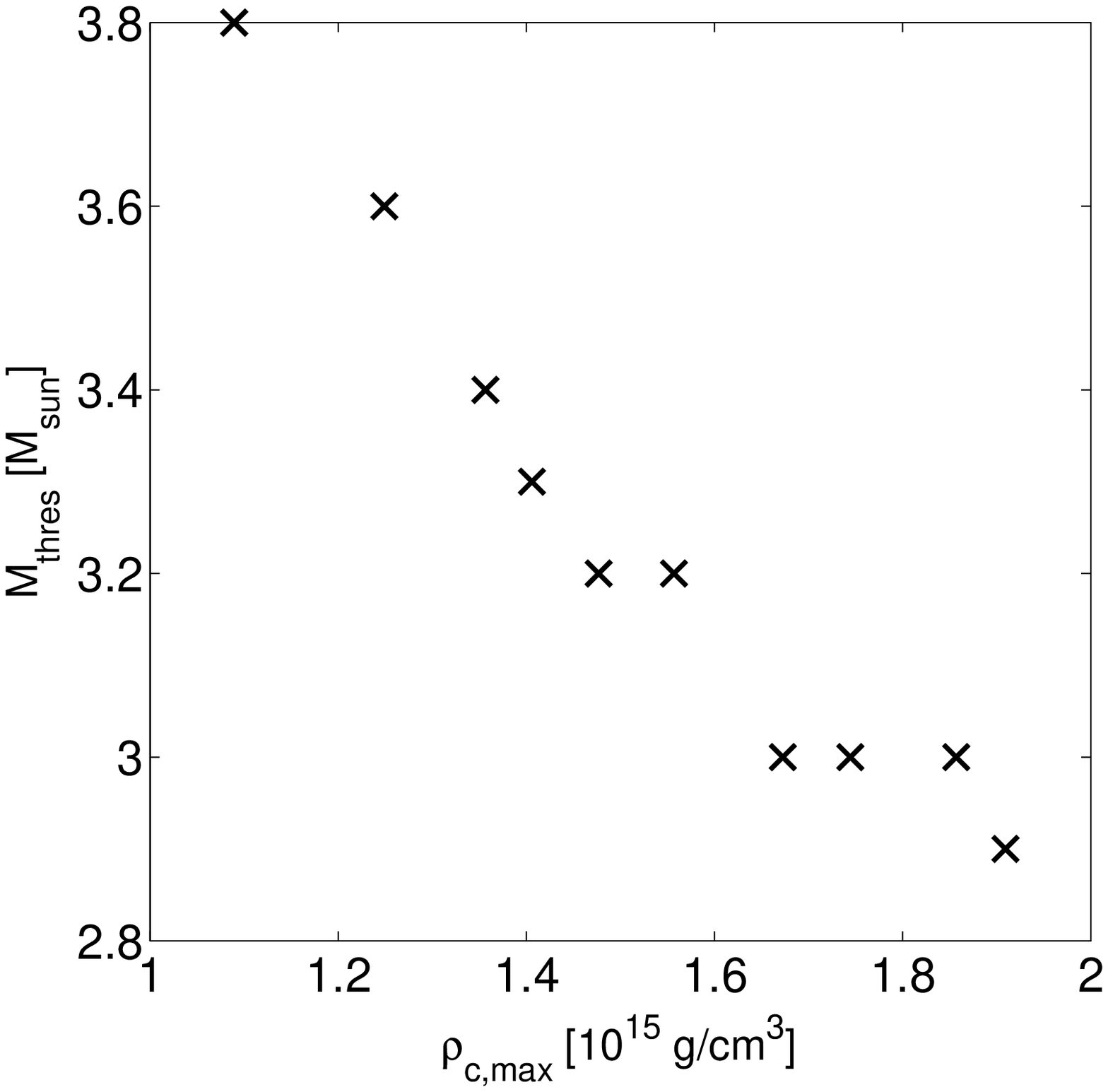}
\caption{\label{fig:mthres}Threshold NS binary mass distinguishing the prompt collapse to a BH  from the formation of an (at least transiently) stable merger remnant as a function of the maximum central energy density $e_{\mathrm{c,max}}$ of the maximum-mass configuration of cold, nonrotating NSs, for different EoSs.}
\end{figure}

We point out that the relation between $e_\mathrm{c,max}$ and $f_\mathrm{thres}$ can be used to deduce a strict upper limit on the maximum mass. It has been shown that causality requires
\begin{equation}\label{eq:causal}
M_\mathrm{max}\le \sqrt{1.358\times 10^{16}\mathrm{g/cm^3}/e_\mathrm{c,max}}~M_\odot,
\end{equation}
which was derived in~\cite{2010arXiv1012.3208L} considering the causal-limit EoS of~\cite{1997ApJ...488..799K}.
An estimate of $f_{\mathrm{peak}}^{\mathrm{thres}}$ by our extrapolation procedure thus provides a proxy for $e_\mathrm{c,max}$ and an upper bound on $M_\mathrm{max}$ via Eq.~\eqref{eq:causal}. Table~\ref{tab:causal} shows that, in particular for stiff EoSs with high $M_\mathrm{max}$, the estimated limits represent strong constraints, which are only a few per cent above the actual value. The importance of these limits lies in the fact that they are independent of the empirical relation connecting the threshold mass and TOV properties (Eq.~\eqref{eq:mthr}) but instead rely only on an estimate of $f_{\mathrm{peak}}^{\mathrm{thres}}$ and the relation shown in Fig.~\ref{fig:emax}.

\begin{table}
\caption{\label{tab:causal}Upper limits on the maximum mass for different EoSs. $M_{\mathrm{max}}$ and $\hat{M}_{\mathrm{max}}$ are the actual values and the estimate via our extrapolation procedure, respectivley. $\hat{M}_\mathrm{max}^{\mathrm{upper}}$ represents a strict upper limit on the maximum mass established by the estimate of $e_\mathrm{c,max}$ only via Eq.~\eqref{eq:causal}.}
\begin{ruledtabular}
 \begin{tabular}{|l|l|l|l|}
\hline    
EoS   &  $M_\mathrm{max}~(M_\odot)$ & $\hat{M}_\mathrm{max}~(M_\odot)$  & $\hat{M}_\mathrm{max}^{\mathrm{upper}}~(M_\odot)$  \\ \hline
NL3   &  2.79  &  2.68  &  2.85 \\
LS375 &  2.71  &  2.69  &  2.80 \\
DD2   &  2.42  &  2.40  &  2.73 \\
TM1   &  2.21  &  2.28  &  2.76 \\
SFHX  &  2.13  &  2.19  &  2.41 \\
GS2   &  2.09  &  2.07  &  2.41 \\
SFHO  &  2.06  &  1.97  &  2.28 \\
LS220 &  2.04  &  1.98  &  2.37 \\
TMA   &  2.02  &  2.12  &  2.66 \\
IUF   &  1.95  &  2.05  &  2.41 \\
\end{tabular}
 \end{ruledtabular} 
\end{table}

Estimates for the central rest-mass density can be obtained by the fit
\begin{equation} \label{eq:fitrho}
\rho_{\mathrm{c,max}}= 0.828\cdot f_{\mathrm{peak}}^{\mathrm{thres}} -1.130 ~(\pm 0.1)
\end{equation}
with the same units for quantities as in Eq.~\eqref{eq:fite}. The fit is based on the data shown in Fig.~\ref{fig:rhomax}. The maximum deviation of the estimated central rest-mass density from its actual value is below 5\%\ (expect for the NL3 EoS, which shows a deviation of 14\%) (see Table~\ref{tab1}). In parentheses, we provide the modifications to Eq.~\eqref{eq:fitrho} for obtaining upper and lower limits on the central rest-mass density, using the extrapolation method.

It is important to note that the relation between $f_{\mathrm{peak}}^{\mathrm{thres}}$ and $M_{\mathrm{thres}}$ means that $R_{\mathrm{max}}$ or the maximum central density also relate to $M_{\mathrm{thres}}$. This is illustrated in Fig.~\ref{fig:mthres} for the rest-mass density (the corresponding plot for the energy density is very similar). The relation implies that not only $R_{\mathrm{max}}$ but also $\rho_{\mathrm{c,max}}$ can be estimated or constrained from any determination or limit on $M_{\mathrm{thres}}$. This is important, because a bound on the threshold mass may be deduced from any observational identification of a prompt collapse event for high-mass binaries. Also, any identification of a delayed collapse immediately implies a corresponding lower limit on the threshold mass. Observationally, such cases might be distinguished by their electromagnetic counterparts~\cite{1998ApJ...507L..59L,2005astro.ph.10256K},
which are expected to be much weaker in cases of prompt collapse to BH \cite{2013ApJ...773...78B}. The involved binary masses, which will set the limit on the threshold mass, can be inferred from the GW inspiral signal.
\subsection{Further considerations}
The details of the extrapolation procedure described above should be considered to be empirically motivated by the behavior of the curves in Fig.~\ref{fig:f2mtot} and the outlying behavior of data points in Fig.~\ref{fig:slope}. We stress that the precision of the procedure does not depend strongly on these particular  choices. Also, the specific forms of the fit formulae do not change the results significantly and might possibly be optimized to yield even better estimates. Note that the uncertainties of the mass estimates are of the order of the numerical determination of the threshold mass, which in this study is achieved only to a certain accuracy. Given a finite sampling of the binary masses, the numerical value of $M_{\mathrm{thres}}$ can only represent a lower bound to the actual value, which, however, lies at most 0.1~$M_{\odot}$ above. Clearly, this numerical inaccuracy is inherent and reflected in the uncertainty of the extrapolation procedure, which thus may be further improved. We also 
expect that the fit formulae may be modified and tuned to even better estimate high-mass NS properties, e.g. for specific mass regimes. Enlarging the set of EoSs will  also be a good test for the accuracy of our method.  

Recently, two new temperature-dependent EoSs have become available, which include a phase transition to hyperonic matter~\cite{2014arXiv1404.6173B} (the $\mathrm{BHB\Lambda}$ EoS has $M_{\mathrm{max}}=1.95~M_{\odot}$ and $R_{\mathrm{max}}=11.75$~km, while the $\mathrm{BHB\Lambda\phi}$ model leads to $M_{\mathrm{max}}=2.10~M_{\odot}$ and $R_{\mathrm{max}}=11.30$~km). We utilize these models as important test cases to assess the applicability of our method to ``unknown'' EoSs which were not included in the determination of the empirical relations employed by our extrapolation procedure (Eqs.~\eqref{eq:stab} to~\eqref{eq:fitrho}). These EoSs are a particular challenge because they involve phase transitions and thus represent rather extreme cases. Using GW data of low-mass binaries, the properties of the maximum mass configurations are recovered by our extrapolation procedure within the same error bars (of a few per cent) as for the EoS sample introduced in Sect.~\ref{sec:sim}. In App.~\ref{app:exeos} we estimate how a larger set of EoSs including also very extreme cases might affect the anticipated error bars under the pessimistic assumption that no further refinement of our method is possible.

The final accurcay of the extrapolation procedure depends on the errors of the slope and of the frequency determination. The error of the slope will be affected by the uncertainty of the individual frequency measurements~\cite{2012PhRvL.108a1101B,2012PhRvD..86f3001B}, by the number of measurements, by the EoS, and in particular by the uncertainty of the values and the exact separation of the distinct binary masses for which GWs are detected. Considering, for example, only two detection events with $M_{\mathrm{tot}}=2.4~M_{\odot}$ and  $M_{\mathrm{tot}}=2.8~M_{\odot}$ one can infer the error on the slope determination. If we assume that the peak frequencies at both binary masses can be measured with a precision of 10~Hz~\cite{Clark2014}, we can quantify the expected error on the intersection of the extrapolating curve with the stability line. We find errors in $M_{\mathrm{thres}}$ above and below one per cent strongly sensitive to the EoS and correspondingly  the proximity of the detection to the stability line. It is important to stress that the extrapolation scheme becomes more accurate if the slope is determined at even higher binary masses.

In the present study, we do not investigate unequal mass binaries, but we note that for fairly unequal 1.2-1.5~$M_{\odot}$ systems the peak frequency is \textit{at most 90~Hz smaller} than the dominant GW frequency of the corresponding equal-mass merger of the same total binary mass. Known NS binaries show smaller mass inequalities (see e.g.~\cite{2012ARNPS..62..485L} for a compilation of the measured masses), and we thus argue that the most likely GW observations will have smaller deviations from the  equal-mass case than the mentioned example. Moreover, we expect that the impact of the mass ratio can be taken into account, once the full dependence of $f_{\mathrm{peak}}$ on the mass inequality and on the EoS is worked out, which we leave for future work.

Finally, we stress that the viability of our procedure to estimate
$M_\mathrm{max}$ and $R_\mathrm{max}$ does {\em not} depend on whether $M_\mathrm{thres}$ is an
accurate measure of the threshold mass for prompt collapse or not.
Equations~\eqref{eq:stab},~\eqref{eq:mthr},~\eqref{eq:rmax}-\eqref{eq:fitrho} could be considered as empirically (i.e., numerically) established relations, whose applicability
does not depend on the exact meaning of the ``stability line''
$f_\mathrm{peak}^\mathrm{thres}(M_\mathrm{thresh})$, i.e., it does not require that this line
defines the boundary for BH formation. However, the procedure relies on the accurate determination of the peak frequency and its slope at $M_\mathrm{tot}\sim 2.7~M_{\odot}$.

\section{Summary and outlook}\label{sec:sum}
Starting from the observation that for a given EoS the dominant GW frequency of the postmerger phase is an increasing function of the binary mass, we construct an extrapolation procedure, which estimates the threshold mass to BH formation from GW observations of low-mass binary NS mergers.  In turn, the threshold mass for prompt collapse can be converted to an estimate of the maximum mass of cold, nonrotating NSs~\cite{2013PhRvL.111m1101B}. We find that two (or more) postmerger GW measurements of binaries with masses of about 2.7~$M_{\odot}$ can be used to infer the maximum mass of nonrotating NSs within a few per cent.
In addition, the estimate of the dominant GW oscillation frequency for a binary at the threshold mass constrains the radius of the maximum-mass configuration of nonrotating NSs with an accuracy of a few per cent. Combining this with the determination of the radius of cold, nonrotating neutron stars of $1.6~M_{\odot}$ (to within a few percent) from the measured GW peak frequencies, allows for a clear distinction of a particular candidate equation of state among a large set of other candidates. Our method is particularly appealing because it reveals simultaneously the moderate and very high-density part of the EoS.

Moreover, our method yields estimates for the maximum central energy density and the maximum central rest-mass density of NSs. Hence, it sets a limit on the highest possible density which can stably exist at the center of relativistic  stars. These estimates may also serve to establish an additional upper limit on the maximum mass which is required by causality and which is independent of the conversion of the threshold mass to the maximum mass.

It is important to stress that the procedure outlined in this work will become more accurate if GW detections become available for binary masses higher than the 2.7~$M_{\odot}$ that were adopted in this study. The closer the final remnant is to the threshold mass, the better will be the accuracy of the our extrapolation procedure.

We also point out that the maximum central density scales with the threshold mass distinguishing the prompt merger collapse from the formation of a NS remnant. This relation implies that any identification of a prompt collapse, e.g. by electromagnetic observations, imposes a limit on the maximum density, apart from constraints on the maximum mass and the radius of the maximum-mass configuration of nonrotating NSs.

The merger models may be improved by taking into account other effects, such as neutrino emission and magnetic fields. Also, the impact of the initial stellar rotation should be explored~\cite{2013arXiv1311.4443B}. The values for $M_{\mathrm{thres}}$ used here represent lower limits because of the finite sampling of the binary parameter space, but are within at most 0.1 $M_\odot$ of the actual values. Hence, the threshold mass may be determined somewhat more accurately by a finer grid of sampled binary parameters. A better determination of the threshold mass may  reduce the (already very small) uncertainties of the extrapolation method proposed here. 

Extending the set of microphysical EoSs beyond what was considered here will refine our method. A different, larger set of EoSs may slightly shift the empirical relations between various quantities that we constructed.  One may find slightly better descriptions for certain regimes or for a specific quantity under consideration. For more than two measurements one should consider the possibility of a higher-order extrapolation. Also, the effects of unequal-mass binaries should be investigated in more detail. Finally, it will be crucial to explore the detectability and the expected observational error bars of the postmerger GW emission (see~\cite{2013arXiv1312.1862M,Clark2014}), in order to determine the overall uncertainty of the procedure in experimental applications.

\appendix
\section{Impact of an extended EoS sample} \label{app:exeos} 
\begin{figure}
\includegraphics[width=8.9cm]{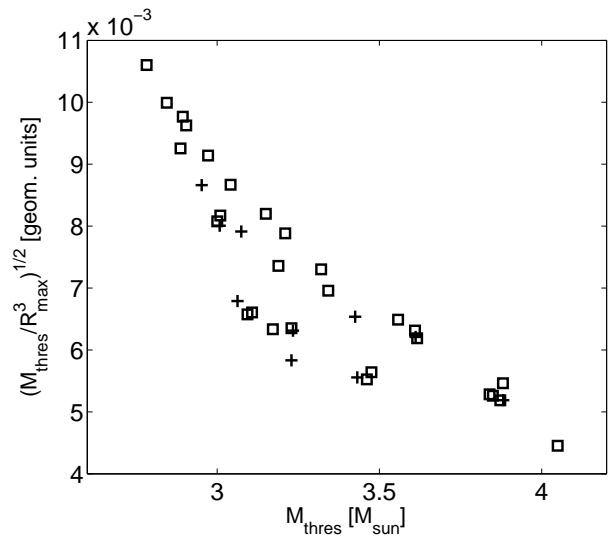}
\caption{\label{fig:alleos}Approximate dominant GW frequency $f_{\mathrm{peak}}^{\mathrm{thres}}$ of the most massive NS merger remnant as a function of the corresponding estimated total binary mass $M_{\mathrm{thres}}$ for a large sample of EoSs. The estimates are entirely based on the stellar properties of cold, nonrotating NSs (see text). The plus signs represent the estimated properties for the set of EoSs employed in this work. The squares display the estimated threshold properties for a larger set of EoSs, which covers also the extreme cases.}
\end{figure}
Given the still limited set of available temperature-dependent EoSs, at this stage we cannot fully exclude that a larger sample of EoSs may to some extent increase the error bars of our method or require a certain refinement of our procedure. To address this issue we note that $f_{\mathrm{peak}}^\mathrm{thres}$ scales approximately with $\sqrt{M_{\mathrm{thres}}/R_{\mathrm{max}}^3}$, while the threshold mass $M_{\mathrm{thres}}$ is approximately given by $M_{\mathrm{thres}}=(-3.38\cdot C_{\mathrm{max}}+2.43)\cdot M_{\mathrm{max}}$ with $C_\mathrm{max}$ being the compactness $G M_\mathrm{max}/(c^2R_\mathrm{max})$ of the maximum-mass configuration~\cite{2013PhRvL.111m1101B}. Hence, both $f_{\mathrm{peak}}^\mathrm{thres}$ and $M_{\mathrm{thres}}$ can be approximately determined by the stellar properties of cold, nonrotating NSs, namely, $M_{\mathrm{max}}$ and $R_{\mathrm{max}}$. In Fig.~\ref{fig:alleos} we show the stability line (similar to Fig.~\ref{fig:fstabmstab}) computed by means of the approximate expressions for $f_{\mathrm{peak}}^\mathrm{thres}$ and $M_{\mathrm{thres}}$ employing $M_{\mathrm{max}}$ and $R_{\mathrm{max}}$ from a larger set of EoSs. Apart from the estimated values for the temperature-dependent microphysical EoSs considered in this paper (plus signs), we also display the approximate threshold properties for the EoSs employed in~\cite{2012PhRvD..86f3001B}, which cover the full range including in particular the extreme cases (see Fig.~4 in~\cite{2012PhRvD..86f3001B}). It is apparent from Fig.~\ref{fig:alleos} that the larger sample leads only to a small broadening of the relation at intermediate threshold masses, where our method achieves a good accuracy.

From Fig.~\ref{fig:alleos} one may conclude that compared to the stability limit found in this work (represented by the plus signs) a new stability limit fitted to a larger set of EoSs (all datapoints in Fig.~\ref{fig:alleos}) may be shifted by about 0.05~$M_\odot$ to higher $M_{\mathrm{thres}}$ compared to the original line. Shifting the stability limit (Eq.~\eqref{eq:stab}) by 0.05~$M_\odot$ to higher $M_{\mathrm{thres}}$, our procedure estimates $M_{\mathrm{max}}$ from the low-mass binary GW signals with a maximum deviation from the actual values of at most 0.15~$M_\odot$. For most of the available models the maximum mass is overestimated. Since some of the available models are located at the extreme left with respect to the new stability (plus signs in Fig.~\ref{fig:alleos}), the resulting error bar of 0.15~$M_\odot$ should be considered to be the maximum possible deviation. Note, however, that some kind of tuning of the procedure for a new function describing the stability limit is likely to reduce this error bar substantially. Employing the modified stability limit for the estimate of $R_\mathrm{max}$ yields essentially the same deviations from the actual values as our original procedure (650~m for the NL3 EoS, while for all other models $R_\mathrm{max}$ is recovered to within 420~m or even better). These estimates show that even for the pessimistic case that no further refinements of our method are applied, two EoSs like in Fig.~\ref{fig:MR2} are clearly distinguishable by means of our extrapolation procedure. The precision of the radius determinations remains practically unchanged, while only the error bars on the maximum mass are somewhat enlarged.

We point out that a possible refinement of our method could consist of describing the stability line as a band embracing all models. For the limited set of EoSs explored in this paper we notice that models which have a relatively steep slope in $f_\mathrm{peak}(M_\mathrm{tot})$ yield $f_\mathrm{peak}(M_\mathrm{tot})$ relations which tend to terminate at the lower left edge of the band. For EoSs which result in a relatively flat slope the $f_\mathrm{peak}(M_\mathrm{tot})$ relations extend until the right edge of the band describing the stability line (see the discussion in Sects.~\ref{sec:idea} and~\ref{sec:extra}). Thus the slope $f_\mathrm{peak}(M_\mathrm{tot})$ may serve to pick an appropriate description of the stability line.

Finally, we note that the approximate scaling relations for $f_{\mathrm{peak}}^\mathrm{thres}$ and $M_{\mathrm{thres}}$ tentatively explain the flattening of the stability line at higher threshold masses (see Fig.~\ref{fig:fstabmstab}; also visible in Fig.~\ref{fig:alleos}). To first order the relation $M_\mathrm{max}(R_\mathrm{max})$ for different EoSs may be approximated by a linear function or a mildly convex function (see Fig.~\ref{fig:MRpoints} or Fig.~4 in~\cite{2012PhRvD..86f3001B}). Inserting a linear or convex function $M_\mathrm{max}(R_\mathrm{max})$ into the approximate formulae for the threshold properties yields a concave stability line.

\begin{acknowledgments}
We are grateful to J. Lattimer for helpful discussions. We thank M. Hempel for providing EoS tables. A.B. is a Marie Curie Intra-European Fellow within the 7th European Community Framework Programme (IEF 331873). This work was supported by the Deutsche Forschungsgemeinschaft through Sonderforschungsbereich Transregio 7 ``Gravitational Wave Astronomy'', and the Cluster of Excellence EXC 153 ``Origin and Structure of the Universe''. Partial support comes from ``NewCompStar'', COST Action MP1304. The computations were performed at the Rechenzentrum Garching of the Max-Planck-Gesellschaft, the Max Planck Institute for Astrophysics, and the Cyprus Institute under the LinkSCEEM/Cy-Tera project.
\end{acknowledgments}

%\bibliography{references}

\end{document}